\documentclass{iau} 
\usepackage{graphicx}

\usepackage{caption}
\title[$\rm{K4000}$ in the spectra of blazar candidates] %% give here short title %%
{Determination of \textbf{$\boldsymbol{\rm{K4000}}$} of potential blazar
candidates among EGRET unidentified gamma-ray sources}

\author[Emmanuel Uwitonze, Pheneas Nkundabakura \& Tom Mutabazi]   %% give here short author list %%
{Emmanuel Uwitonze$^1$, Pheneas Nkundabakura$^2$,
 \and Tom Mutabazi$^1$}
\affiliation{$^1$Physics Department, Mbarara University of Science and Technology, P.O. Box 1410, Mbarara, Uganda\\
email: uwitonze\_emmanuel@yahoo.com\\
$^2$Department of Mathematics, Sciences and Physical Education, University of Rwanda-College of Education, P.O. Box
5039, Kigali, Rwanda}

\pubyear{2019}
\volume{356}  %% insert here IAU Symposium No.
\jname{Nuclear Activity in Galaxies Across Cosmic Time}
\editors{M.\,Povi\'c, P.\,Marziani, J.\,Masegosa, H.\,Netzer,\\ S.\,H.\,Negu, \& S.\,B.\,Tessema, eds.}
\begin{document}

\maketitle

\begin{abstract}
Blazars are radio-loud Active Galactic Nuclei (AGN) with relativistic jets oriented towards the observer's
line-of-sight. Based on their optical spectra, blazars may
be classified as flat-spectrum radio quasars (FSRQs) or
BL Lacs. FSRQs are more luminous blazars with both narrow and broad emission and absorption lines, while BL
Lacs are less luminous and featureless. Recent studies show that blazars dominate ($\sim93$\,\%)
the already-identified EGRET sources (142), suggesting that among the unidentified sources (129) there could still be faint blazars. Due to the presence of a strong
non-thermal component inside their jets, blazars are found to display a weaker depression at $\sim 4000$\,\AA\, ($\rm{K4000} \leq0.4$). 
In this study, we aimed at determining the $\rm{K4000}$ break for a 
selected sample among the potential blazar candidates from unidentified EGRET 
 sources to confirm their blazar nature. We used two blazar candidates, 3EG J1800-0146 and 3EG J1709-0817
  associated with radio counterparts, J1802-0207 and J1713-0817, respectively. Their optical counterparts were
  obtained through 
 spectroscopic observations using Robert Stobie spectrograph (RSS) at the Southern African Large Telescope (SALT) in South Africa.
 The observed Ca II H\,$\&$\,K lines depression at $\sim4000$\,\AA\, in spectra of these sources 
 show a shallow depression, $\rm{K4000}=0.35\pm0.02$ and $0.24\pm0.01$, respectively, suggesting that these sources are blazar candidates.
 Moreover, the redshifts $\rm{z}=0.165$ and $0.26$ measured in their spectra confirm the extragalactic nature of these sources.

\end{abstract}

\firstsection % if your document starts with a section,
\section{Introduction}
Different studies reveal that out of 271 sources detected by the
Energetic Gamma-Ray Experiment Telescope (EGRET) during its mission, $\sim142$ sources were identified and $\sim93$\,\% of them 
were associated with blazars (e.g.,  \cite[Hartman et al. 1999]{Hartman_etal99}; \cite[Sowards-Emmerd et al. 2003]
{Sowards-Emmerd_etal03}; {\cite[Nkundabakura \& Meintjes 2012]{NkundabakuraMeintjes12}). It is assumed that among 129 ($\sim\,48$\,\%) remaining
unidentified EGRET sources, there could be still faint blazars in abundance.
There have been efforts to try to identify blazars from the unidentified EGRET sources via
their variability and spectral energy distribution (SED). This study
joined this effort and aimed at determining the extent of non-thermal emissions from two potential blazar candidates selected 
among the remaining unidentified EGRET sources through the analysis of the strength of their
Ca II H\,$\&$\,K break at $\sim$\,4000\,\AA\,($\rm{K4000}$). Blazars are a special class of AGN with
a weaker Ca II contrast, i.e., $\rm{K4000} \leq0.4$ (\cite[Marcha et al. 1996]{Marcha_etal96}; \cite[Caccianiga
et al. 1999]{Caccianiga_etal99};
\cite[Landt et al. 2002]{Landt_etal02}), 
due to the contribution of the nuclear non-thermal component produced inside their jets, i.e., for blazars, this Ca II H\,$\&$\,K break decreases as the non-thermal emission
 from jet increases and vice-versa. In fact, observations in optical, radio and X-rays show that the increase
 in the jet's luminosity leads to  the decrease in the Ca II H\,$\&$\,K break (\cite[Landt et al. 2002]{Landt_etal02}).

 \section{Data source and methods}
\label{data and methods}
The study used two blazar candidates (3EG J1800-0146 and 3EG J1709-0817) from a sample of 13 blazar candidates
selected from {\cite[Nkundabakura \& Meintjes (2012)]{NkundabakuraMeintjes12} which
were selected from the 3EG catalogue. The spectroscopic raw data were obtained at 
the Southern African Large Telescope (SALT) with RSS spectrograph. The raw data of the two sources were reduced using the
Image Reduction and Analysis Facility (IRAF) data reduction pipeline. The presence of nuclear non-thermal component in the host galaxy of each 
source was confirmed by measuring the $\rm{K4000}$ in the rest frame. For each spectrum, the contrast
was calculated using the relation: 
 \begin{equation}
  \rm{K4000=\frac{f^+-f^-}{f^+}},
  \label{eqn1}
 \end{equation}
where $\rm{f^{-}}$ and $\rm{f^{+}}$ are average fluxes within 3750\,\AA\,--3950\,\AA\, and 4050\,\AA--4250\,\AA\,,
respectively, in rest frame (\cite[Caccianiga et al. 1999]{Caccianiga_etal99};
\cite[Landt et al. 2002]{Landt_etal02}). The fluxes measured in each spectrum are shown in Table \ref{D4000}.
\section{The spectra of 3EG J1800-0146 and 3EG J1709-0817}
\label{Results}
Figure \ref{spectra} (left panel) represents the optical spectrum of 3EG J1800-0146. The spectrum
features some absorption lines at redshift $\rm{z}=0.165$, e.g.,
Ca II H\,\&\,K ${\lambda}{\lambda}$3969\,\AA, 3934\,\AA\,, Mg\,I\,b 
${\lambda}{\lambda}$5169, 5175, 5184\,\AA\, and Na\,D ${\lambda}{\lambda}$ 5890, 5896\,\AA\, in the rest frame. $\rm{H_{2}0}$ absorption 
(7168\,--\,7394\,\AA) from the Earth's atmosphere, not redshifted has been detected in this source. The line
centre, continuum, flux and the 
    equivalent width ($\rm{W_{\lambda}}$) shown in Table \ref{tab:table2} were 
    measured using SPLOT command in IRAF.
    \begin{table}
 \small
 %\addtolength{\tabcolsep}{-2pt}
  \centering
    \caption{Absorption lines in 3EG J1800-0146 and 3EG J1709-0817. 
    Each line is redshifted by $\rm{z}=0.165$ and 0.26, respectively.}
    \label{tab:table2}
    \begin{tabular}{lc lc lc lc lc}
    \hline
    Absorption line & Line centre& Continuum&Flux&$W_{\lambda}$\\
                      &(\AA)&($\rm{erg ~s^{-1} {cm}^{-2}} {\textup{\AA}}^{-1}$)&(erg $\rm{s^{-1}} \rm{{cm}^{-2}}{\textup{\AA}}^{-1}$)&(\AA)
                      \\
                      \hline 
       &   &3EG J1800-0146&&\\\\
      Ca II K&4580.18&2.55${\times}10^{-16}$ &4.95${\times}{10}^{-16}$&1.93\\
      Ca II H&4625.53&2.30${\times}10^{-16}$&8.00${\times}{10}^{-17}$&0.34\\
      Mg I b&6046.36&1.25${\times}10^{-15}$&8.25${\times}{10}^{-16}$&0.65\\
      Na D&6863.19&1.81${\times}10^{-15}$&3.25${\times}{10}^{-15}$&1.79\\\\
      &   &3EG J1709-0817&&\\\\
      Ca II K&4973.92&3.483${\times}10^{-16}$ &6.95${\times}{10}^{-16}$&1.996\\
      Ca II H&4997.05&3.472${\times}10^{-16}$&6.69${\times}{10}^{-16}$&1.927\\
      Fe d&5882.79&8.13${\times}10^{-16}$&2.03${\times}{10}^{-15}$&2.49\\
      \hline 
    \end{tabular}
  \end{table}
  
  \begin{figure}[!h]
 \begin{tabular}{c c}
 %\subfigure[Optical spectrum of 3EG J1800-0146  ]{
\includegraphics[width=2.54in]{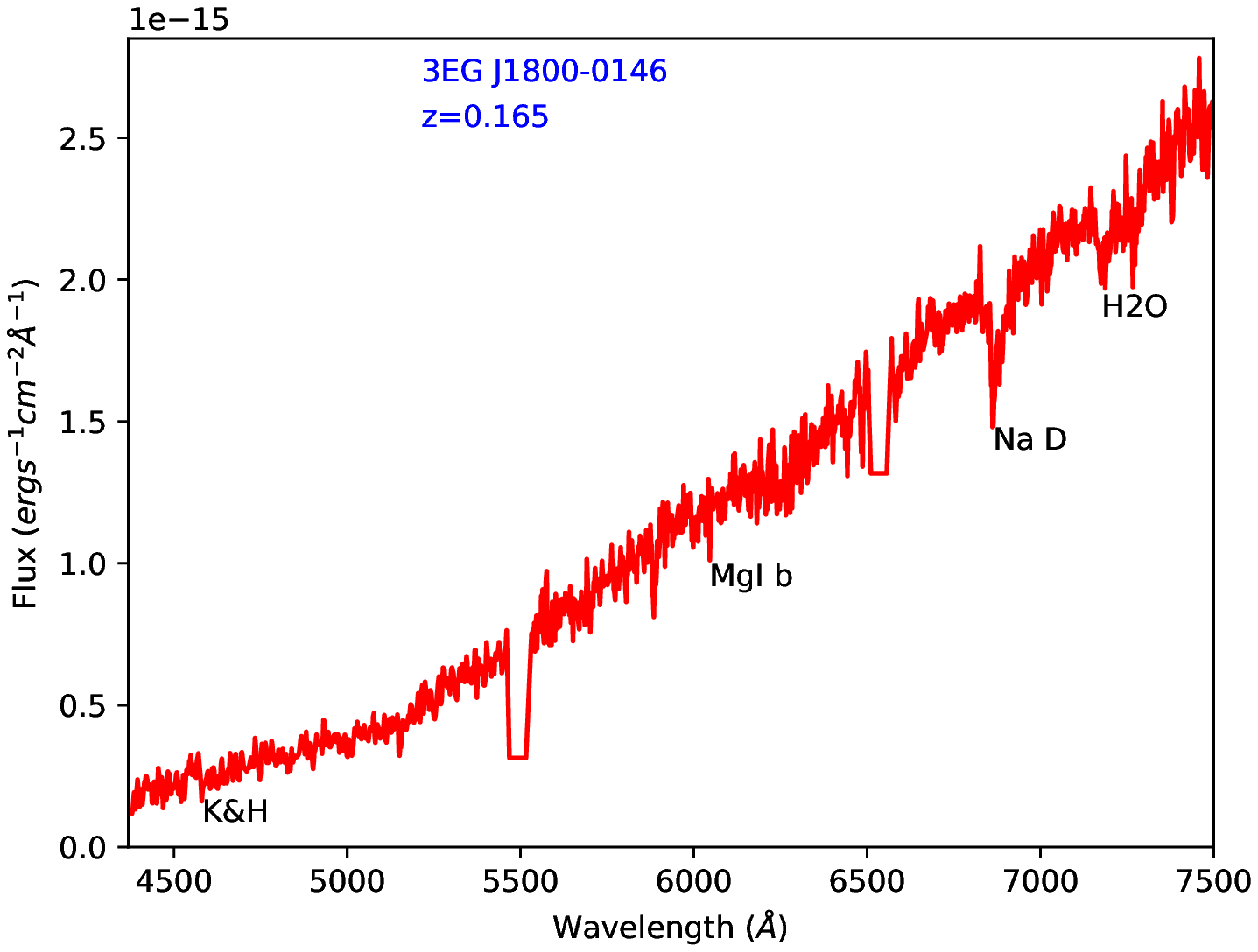}
 %}
%\subfigure[Optical spectrum of 3EG J1709-0817]{
\includegraphics[width=2.54in]{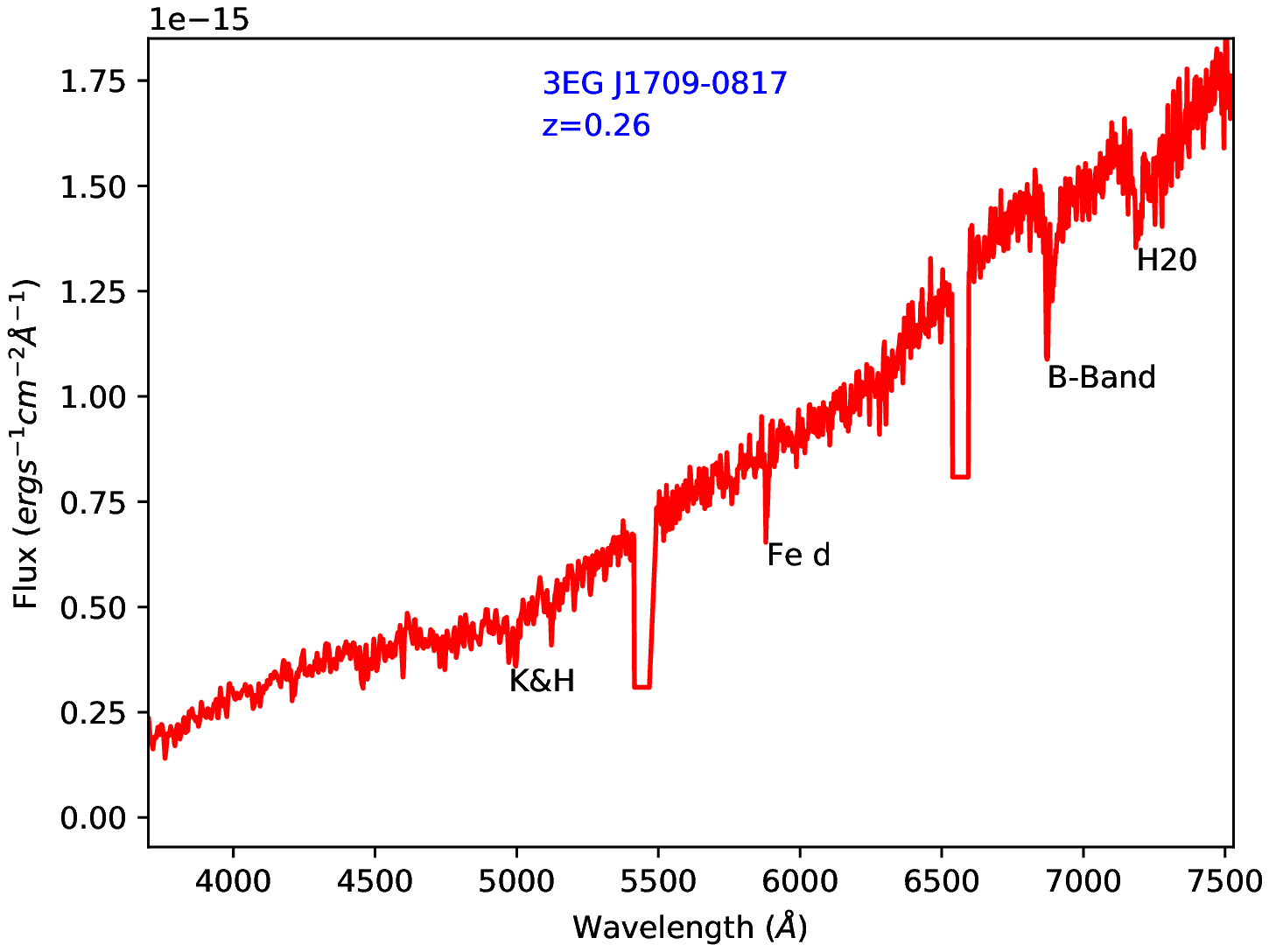}
%}
\end{tabular}
\caption{The left panel shows the spectrum of 3EG J1800-0146. The right panel shows the spectrum of 3EG J1709-0817.}

\label{spectra}
\end{figure}
It shows a shallow depression ($\rm{K4000}=0.35\pm0.02$), suggesting the
presence of a strong non-thermal emission in this source, emanating
from a jet oriented at a small angle. The spectrum of this source resembles that of
a BL Lac (\cite[Marcha et al. 1996]{Marcha_etal96}; \cite[Landt et al. 2002]{Landt_etal02}). On the other hand,
Figure \ref{spectra} (right panel) shows
the spectrum of 3EG J1709-0817. It features some absorption lines at redshift $\rm{z}=0.26$,
e.g., Ca II H\,\&\,K ${\lambda}{\lambda}$3969\,\AA, 3934\,\AA\, and Fe\,d at $\lambda$4668\,\AA\, in the rest frame (Table \ref{tab:table2}). The telluric absorption lines
resulting from Earth's atmosphere, not redshifted have been identified in this
object, e.g., $\rm{H_{2}}O$ (7168\,--\,7394\,\AA), and B-Band (6867\,--\,6884\,\AA). The spectrum displays
a shallow depression, $\rm{K4000}=0.24\pm0.01$. 
 This is within the range of $\rm{0.25\leq K4000\leq0.4}$ of typical BL Lac candidates
 (\cite[Marcha et al. 1996]{Marcha_etal96}). In Table \ref{D4000}, we present the average
 fluxes, $\rm{f^{-}}$ and $\rm{f^{+}}$ used for the measurement of $\rm{K4000}$ in the two spectra.

\begin{table}
%\scriptsize%{7}{7}
\small
%\addtolength{\tabcolsep}{-8pt}
\centering
    \caption{Ca depression near 4000\,\AA\, in 3EG J1800-0146 and 3EG J1709-0817 spectra.}
    \label{D4000}
    \begin{tabular}{lc lc lc lc lc}
    
    \hline

      Object &Ca band& Ca band& z&  Flux&$\rm{K4000}$&Class by $\rm{K4000}$ \\
             &(\AA, rest frame)&($\textup{\AA}$, redshifted)& &(erg $\rm{{s}^{-1}} \rm{{cm}^{-2}} \textup{\AA}^{-1}$)& & &\\
      \hline 
      %\hline
      
       3EG J1800-0146& 3750\,--\,3950&4368.75\,--\,4601.75&0.165& $\rm{f^{-}=(2.170\pm0.040){\times}{10}^{-16}}$& &\\
      
                    &4050\,--\,4250&4718.25\,--\,4951.25& & $\rm{f^{+}=(3.322\pm0.035){\times}{10}^{-16}}$&0.346$\pm$0.02& BL Lac\\   
      3EG J1709-0817& 3750\,--\,3950&4725\,--\,4977&0.26& $\rm{f^{-}=(4.333\pm0.030){\times}{10}^{-16}}$&\\
                    &4050\,--\,4250 &5103\,--\,5355& &$\rm{f^{+}=(5.720\pm0.050){\times}{10}^{-16}}$&0.242$\pm$0.01& BL Lac&\\

      \hline
    \end{tabular}
  \end{table}
%  \begin{figure}[h!]
%   
%  % \vspace*{-2.0 cm}
% 
%  \begin{center}
%  \includegraphics[width=3.4in]{3EGJ1709-0817_ap6_final_spectrum.eps}
%  % \vspace*{-1.0 cm}
% \caption{Optical spectrum of 3EG J1709-0817.}
% \label{spec0006}
% \end{center}
% \end{figure}

\iffalse
\subsection{3EG J0159-3603}
The spectrum of 3EG J0159-3603 (Figure \ref{spec0002}) displays a deeper depression $\rm{K4000}=0.54\pm0.03$.
This is not in agreement with the depression expected for a non-thermal source, implying that this source may not a blazar. 
\begin{figure}[h!]
 \begin{center}
 %\vspace*{-2.0 cm}
 \includegraphics[width=3.4in]{3EGJ0159-3603_cut_corrected.eps}
\caption{Optical spectrum of 3EG J0159-3603.}
%\vspace*{-1.0 cm}
\label{spec0002}
\end{center}
\end{figure}

\fi

\section{Conclusions}
\label{conclusions}
From our spectroscopic data, we confirmed 3EG J1800-0146 and 3EG J1709-0817 as
extragalactic objects with redshift $\rm{z=0.165}$ and 0.26, respectively, and as BL Lacs
through the strength of their $\rm{K4000}$ contrast.

%\section{Acknowledgement}
%The authors are thankful to the IAU S356 Symposium organisers for the invitations to present this at the conference.
% The financial support from the SIDA through ISP to the East African Astrophysics Research Network. 


\begin{thebibliography}{}

% \bibitem[Caccianiga \etal\ (1999)]{Caccianiga_etal99}
% {Caccianiga, A., Maccacaro, T., Wolter, A., Della Ceca, R., \& Gioia I.M,} 1999,
%  \textit{ApJ}, 513, 51-68
 
 
 \bibitem[Caccianiga \etal\ (1999)]{Caccianiga_etal99}
{Caccianiga, A., Maccacaro, T., Wolter, et al.,} 1999,
 \textit{ApJ}, 513, 51
 
 
 \bibitem[Hartman \etal\ (1999)]{Hartman_etal99}
 {Hartman, C., Bertsch, L., Bloom, et al.,} 1999,
\textit{ApJS}, 123, 79

%  \bibitem[Marcha \etal\ (1996)]{Marcha_etal96}
%  {Marcha, M., Browne, A., Impey, E., \& Smith P.S.,} 1996,
% \textit{MNRAS}, 281, 425-448



%  \bibitem[Landt \etal\ (2002)]{Landt_etal02}
%  {Landt, H., Padovani, P., \& Giommi P., } 2002,
%  \textit{MNRAS}, 336, 945-956
 
 
 \bibitem[Landt \etal\ (2002)]{Landt_etal02}
 {Landt, H., Padovani, P., Giommi, et al., } 2002,
 \textit{MNRAS}, 336, 945



\bibitem[Marcha \etal\ (1996)]{Marcha_etal96}
 {Marcha, M., Browne, A., Impey, et al.,} 1996,
\textit{MNRAS}, 281, 425
 
 

% 
% \bibitem[\protect\citeauthoryear{Fan}{2005}]{Fan2005}
% Fan., 2005, aap, 436, 799-804

%  \bibitem[Nkundabakura \etal\ (2012)]{Nkundabakura_etal12}
%  {Nkundabakura, P., \& Meintjes P.J.,} 2012,
% \textit{MNRAS}, 427, 859-871


\bibitem[Nkundabakura \etal\ (2012)]{Nkundabakura_etal12}
 {Nkundabakura \& P., Meintjes.,} 2012,
\textit{MNRAS}, 427, 859


% \bibitem[\protect\citeauthoryear{Stocke}{1991}]{Stocke1991}
% Stocke., 1991, apjs, 76, 813-874


 
 
% \bibitem[\protect\citeauthoryear{Dressler}{1987}]{Dressler1987}
% Dessler., 1987, AJ, 94, 899-905
% 
% \bibitem[\protect\citeauthoryear{Browne}{1993}]{Browne1993}
% Browne., 1993, mnras, 261, 795

%  \bibitem[Sowards-Emmerd \etal\ (2003)]{Sowards-Emmerd_etal03}
%  {Sowards-Emmerd, D., Romani, W., \& Michelson P.F.,} 2003,
% \textit{APJ}, 590, 109-122

\bibitem[Sowards-Emmerd \etal\ (2003)]{Sowards-Emmerd_etal03}
 {Sowards-Emmerd, D., Romani, R.~W., Michelson, et al.,} 2003,
\textit{ApJ}, 590, 109





 \end{thebibliography}
\end{document}